\documentclass[twocolumn,showpacs]{revtex4b4}
\usepackage{graphicx}
\usepackage{dcolumn}
\usepackage{amsmath}

\newcommand{\bu}{{\bf u}}

\newcommand{\bF}{{\bf F}}
\newcommand{\br}{{\bf r}}
\newcommand{\bx}{{\bf x}}

\newcommand{\bnab}{{\bf \nabla}}

\begin{document}
\title{Two--point microrheology and the Electrostatic Analogy}
\author{Alex J.~Levine and T.C.~Lubensky}
\affiliation{Department of Physics and Astronomy, University of Pennsylvania,
Philadelphia, PA 19104}

\date{\today}
\begin{abstract}
The recent experiments of Crocker {\it et al.\/} suggest that
microrheological
measurements obtained from the correlated fluctuations of widely-separated
probe particles determine the rheological properties of soft, complex materials
more accurately than do the more traditional particle autocorrelations.
This presents an interesting problem in viscoelastic dynamics.
We develop an important, simplifing analogy between the present viscoelastic
problem and classical electrostatics.  Using this analogy and direct
calculation we analyze both the one and two
particle correlations in a viscoelastic medium in order to explain this
observation.

\end{abstract}
\maketitle
\section{Introduction}

The utility of microrheology in probing the structure of soft materials
has been recognized for some time\cite{Mason:95}.  This experimental
technique
uses the position correlations of thermally fluctuating, rigid
probe particles embedded in a soft medium to measure the response of
those particles to an external force\cite{FDT}.  From that response function one
can determine the rheological properties of the material.  In addition to
this passive form of measurement, the response function can also be obtained
directly by applying an external force to the probe beads.  Such active
versions of the experiment have been performed using magnetic
particles\cite{Magnetic}.  It may be pointed out that sedimentation
experiments used to measure the viscosity of fluids can be thought
of as the zero-frequency limit of the active form of microrheology experiments.

The fundamental assumption underlying the data reduction in these measurements
is the relation between the response function of a bead (rigid, spherical
particle) to an externally
applied force and the rheological properties of the medium
in which that bead is embedded.  The form of the single sphere
(of radius $a$)
response function, $\alpha^{(1,1)}(\omega)$, that is commonly used is
the generalized Stokes--Einstein relation (GSER), which has the form:
\begin{equation}
\label{GSER}
\alpha^{(1,1)}_{ij} = \frac{ 1}{6 \pi a G(\omega)} \delta_{ij},
\end{equation}
where $G(\omega)$ is the complex shear modulus of the medium.
The superscript points out that the we are considering the position
response of a sphere to a force applied to that same sphere. The subscripted
indices are the usual vectorial indices.

This response function owes it name to the fact that, for a Newtonian,
viscous fluid where $G(\omega) = -i \omega \eta$, $\alpha^{(1,1)}$
reduces to the
Stokes mobility of a sphere of radius $a$. A microrheological experiment
in such a one--component Newtonian medium consists of measuring
the position auto-correlations of a sphere diffusing in the Newtonian fluid.
These correlations are controlled by the sphere's diffusivity,
which is obtained from the Stokes mobility via the Einstein relation.
Thus, the
measured position autocorrelations of the sphere
allow one to calculate, using the response function,
Eq.~(\ref{GSER}), the fluid's viscosity.  This
viscosity encodes all the rheology of the Newtonian fluid.

We have already examined the validity of
the generalization of the GSER to a viscoelastic medium in  previous articles
\cite{LevinePRL:2000,Levine:2000} and have found that
in many experimental systems there is
a significant frequency range over which Eq.~(\ref{GSER}) is a
good approximation to the single-sphere response function.
At frequencies where the single particle
response function deviates significantly from  Eq.~(\ref{GSER}),  the
breakdown of the GSER can be attributed to one of two sources:
i) inertial effects
at high frequencies, or  ii) the effective decoupling of network and
fluid dynamics at very low frequencies.
We have found that inertial effects typically become significant at
such high frequencies that we may safely ignore them here. Moreover, in this article,
we will incorporate the appearance of non-shear modes by giving our
course-grained model of a viscoelastic medium a complex,
frequency--dependent bulk modulus in addition to its frequency-dependent,
complex shear modulus.

Nevertheless, there still remain fundamental questions regarding the
interpretation of microrheological data.  In this article we address
one such question: Given that the presence of the probe sphere can
locally perturb the micro-structural and, therefore, the rheological
properties of the medium, how can one extract information about the bulk,
unperturbed medium?  In other words we imagine that each probe sphere
is surrounded by a pocket of perturbed material with rheological
properties diferent from those of the bulk.
For microrheology to be a useful experimental probe, it must be possible
to extract the bulk, unperturbed viscoelastic moduli of the medium
from the measured correlation functions.  However, given that the probe
sphere is coupled to the bulk medium by a pocket of material whose
rheological properties are modified by the introduction of that
particle, one must assume that the correlations actually measure
some convolution of the perturbed and bulk material properties.

The assumption of the presence of such pockets is
quite reasonable in many complex liquids.  The pocket, for example,
may be a result of the equilibrium distribution of polymers near an
impenetrable bead in solution; or it may be the
result of quenched inhomogenities produced by the action of
the probe during the formation of the medium.  For example, in
microrheological studies of polymerized F-actin,  monomeric G-actin
is polymerized in a solution already containing the probe
particles\cite{Schnurr:97}.  The proximity of the
probe particle may locally affect the polymerization kinetics and
lead to a positionally dependent F-actin density near the probe spheres
that is independent of equilibrium effects such as the steric interaction
between the actin rods and the probes.
We do not consider this situation is detail, but later in this article
we do explore the consequences
of polymer depletion near the surface of the bead in equilibrium.
In this example the steric interaction of the polymers with the
probe particle produces
regions surrounding the beads with a softer shear modulus than the bulk.
Recently Crocker {\it et al.\/}\cite{Crocker:2000} have proposed a
modification of the standard microrheological technique that can remove the
effect of the perturbed pockets by studying the {\it inter--}particle
position correlations of rather distant probe spheres.  This
claim can be reexpressed in terms of the two--particle response function or
compliance tensor,
$\alpha^{(n,m)}_{ij}$, defined by
\begin{equation}
\label{response-full}
r^{(n)}_i(\omega) = \alpha^{(n,m)}_{ij}(\br^{(m)} - \br^{(n)}, \omega)  \,
F^{(m)}_j(\omega ),
\end{equation}
where $\br^{(n)}(\omega)$ is the displacement of the $n^{\rm th}$ sphere and
$\bF^{(m)}$ is the external force applied to the $m^{\rm th}$ sphere. The claim
is that when the spheres (of radius $a$) are separated by a distance
$r$, $r \gg a$, $\alpha^{(n,m)}_{ij}(\br, \omega)$ for $n\neq m$  depends
upon only the bulk properties of the material.

In this paper we demonstrate the validity of the Crocker hypothesis by solving
the elastic problem of two spheres embedded in an inhomogeneous elastic medium.
We calculate the mutual response function of these beads, $\alpha_{ij}^{(1,2)}$
 and show, in the limit
mentioned above, that this response function measures the bulk rheological
properties of the medium independently of the rheological properties of
the regions immediately surrounding the two beads.

\begin{figure}[bp]
\scalebox{0.4}{\includegraphics{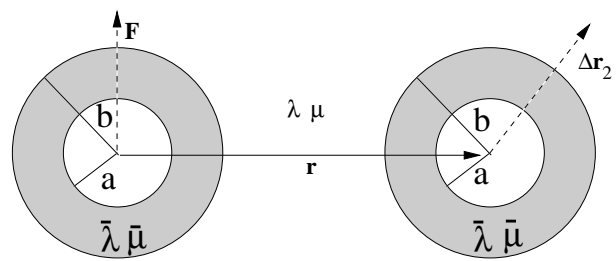}}
\caption{%
Diagram of the simplest inhomogeneous elastic medium consistent with the
assumed rotational symmetry of the problem. Each rigid sphere of radius $a$
is surrounded by a spherical pocket with radius $b$, $b > a$ of material
with elastic constants: $\bar{\lambda}, \bar{\mu}$.  The bulk material has
elastic constants: $\lambda, \mu$.  A force $\protect{\bF}$ is applied to
sphere 1 (on the left). We seek the resulting displacement of sphere 2
(on the right),
$\protect{\Delta \br_2}$.  In the following we will assume that the
separation of the two spheres, $r$ is large compared to $b$; the picture
is not drawn to scale.
}
\label{sphere-pic}
\end{figure}

The remainder of the paper is
organized as follows.  In section \ref{EM} we identify an analogy between
the viscoelastic problem that we posed and the physics of embedded conductors
in an inhomogeneous dielectric.  We use this analogy in combination with
well-known results for the mutual capacitance of two spheres to elucidate
the more complex viscoelastic problem.  This heuristic analogy guides our
approach to the full viscoelastic problem which is studied in
section \ref{visco}.  We approach the full problem in stages by first
considering a rheologically homogeneous material in section \ref{visco-homo}
and then by studying, in section \ref{visco-inhomo}
a simple model of a rheologically inhomogeneous material
consisting of the bulk medium and
``pockets'' of rheologically perturbed material surrounding each probe sphere
as depicted in figure~\ref{sphere-pic}. We show, in the limit that the radii of
these anomalous  pockets are small compared to the separation of the
probe spheres, that the inter--particle response function
can be obtained with a minimum of computational effort through the use of
a global property of the stress tensor.
Most importantly, the leading term
in the inter--particle response function is determined solely by the properties
of the bulk medium.  Following up this result we turn to the more
 computationally complex problem of finding the {\it single\/} particle
response function in this composite medium.  This is accomplished in section
\ref{single-sphere}. This result will be shown to depend on the rheological
properties of both the bulk material and the perturbed material in the
pockets.  It is interesting to note that the combination of
the results of sections \ref{visco-inhomo} and \ref{single-sphere} suggest that
one can experimentally determine the material properties of both the bulk
and perturbed media through a combination of single--particle and two--particle
microrheology experiments.

Motivated by this realization we study in section \ref{differential-shell}
a more physical model of the probe particle in a soft, complex medium.
We now assume that the rheological properties of the medium vary
continuously with the distance from the probe sphere. Our previous algebraic
solution to the two material problem (bulk and perturbed pocket material) now
has to be generalized to a integral technique.
As an example, we apply this technique
to a polymer solution with concentration slightly above $c^{\star}$
in order to study the effect of polymer depletion near the probe sphere.
This problem is revelent to recent experiments on DNA solutions\cite{weeks:00}.
Finally, we summarize these results and conclude in section \ref{final}.

\section{The Electrostatic Analogy}
\label{EM}
To keep our treatment as simple as possible, we will assume that our
viscoelastic medium is characterized by a local relation between the stress
$\sigma_{ij}$ and the strain $u_{ij}$ described by a local,
frequency--dependent,
but possibly spatially varying elastic constant tensor $K_{ijkl}(\bx, \omega)$.
The analysis we present here will have to be modified if the local
stress--strain relation does not hold as is argued to be the case in
systems of experimentaly interest such as actin networks\cite{Maggs}.  Under
our assumptions, the equation of force balance in a linear viscoelastic medium
can be written as
\begin{equation}
\label{basic-elastic}
- \partial_j \left[ ( K_{ijkl}({\bf x},\omega ) \partial_k u_l \right]  =
f_i({\bf x}, \omega),
\end{equation}
where $u_l(\bf x, \omega )$ is the local displacement variable and
$f_i({\bf x}, \omega)$ is a local force density at $\bx$.
We compare the above expression to the
Gauss's law in an inhomogeneous dielectric medium:
\begin{equation}
\label{electric-potential}
-  \partial_j  \left[  \epsilon_{jk}({\bf x}, \omega ) \partial_k
\phi({\bf x}, \omega ) \right]  = 4 \pi \rho({\bf x}, \omega ),
\end{equation}
where $\epsilon_{jk}({\bf x},\omega)$ is the local frequency--dependent
dielectric constant tensor,
$\rho({\bf x}, \omega )$ is the frequency--dependent charge
density and $\phi(\bx, \omega)$ is the electric potential at ${\bf x}$.
In Eq.~(\ref{electric-potential}) we have assumed, as we have done with the
elastic constant tensor, that the dielectric tensor is local.
We consider the above electromagnetic problem at low enough frequencies so that
that we may ignore the transverse electric fields.

Comparing Eqs.~(\ref{basic-elastic}) and (\ref{electric-potential}), we note
that the following correspondence table (see table \ref{table1}) may be drawn:
The charge density in
Eq.~(\ref{electric-potential}) is the scalar analog of the vector source,
$f_i({\bf x},\omega)$
in Eq.~(\ref{basic-elastic}).  Similarly the electric
potential, $\phi(\bx, \omega)$ in Eq.~(\ref{electric-potential}) is analogous
to vector displacement field, $\bu(\bx,\omega)$ in Eq.~(\ref{basic-elastic}),
and  the position--dependent dielectric tensor,
$\epsilon_{ij}({\bf x}, \omega )$, has as its analog in
Eq.~(\ref{basic-elastic}) the elastic constant tensor,
$ K_{ijkl}({\bf x},\omega )$.

\begin{table}
\caption{Correspondence between the electrostatics and viscoelastics
\label{table1}}
\begin{ruledtabular}
\begin{tabular*}{\hsize}{l@{\extracolsep{0ptplus1fil}}c@{\extracolsep{0ptplus1fil}}r}
Electrostatics&Viscoelastics\\
\colrule
potential $\phi ({\bf x})$ & displacement $u_i({\bf x})$\\
charge density $\rho ({\bf x})$ & force density $f_i({\bf x})$\\
dielectric tensor $\epsilon_{ij} ({\bf x},\omega )$& elastic tensor
$K_{ijkl}({\bf x}, \omega )$\\
\end{tabular*}
\end{ruledtabular}
\end{table}
 Finally, we note that the rigidity of objects
embedded in the inhomogeneous viscoelastic medium requires that the
displacement field, $\bu$, be constant on their surfaces.  Therefore, in
order to maintain the analogy between the viscoelastic problem and the
electrostatic problem, we study collections of embedded conducting objects so
that the electric potential is constant on their surfaces.

Recall that the goal of our calculation is to determine the compliance
tensor introduced in Eq.~(\ref{response-full}).  This response function relates
a set of forces applied to rigid objects embedded in an (in general
inhomogeneous) dielectric to the displacements of
those objects.  In order to discuss this calculation in terms of the
simpler electrostatic problem, we need to consider the electrostatic
quantity that is analogous to the compliance tensor. This quantity is the
inverse capacitance tensor of a collection of conducting objects embedded
in an inhomogeneous dielectric.
Since the electrostatic problem is a simpler,
scalar version of the viscoelastic problem, we begin an with analysis of the
that system.
Afterward, insights drawn from the electrostatic problem should lead to
complimentary results in the elastic problem, which remains the actual
problem of interest.

We study a particularly simple realization of the so far arbitrary
inhomogeneous dielectric medium.  The simple model is meant to begin
the study of the ``pocket model'' discussed in the introduction (See figure
\ref{sphere-pic}.) from within the electrostatic analogy.
We consider that the inhomogeneous dielectric is made up of two materials. The
bulk material has dielectric constant
$\epsilon_{ij}(\bx) = \delta_{ij} \epsilon$.  However in concentric pockets
around the conducting spheres (of radius $a$) there are spherical shells of
material ($a < r < b$) with a different dielectric constant:
$\epsilon_{ij}(\bx) = \delta_{ij} \bar{\epsilon}$.  Hereafter we assume
that the dielectric tensor is diagonal and suppress
its tensorial indices.

To support the notion that the off-diagonal elements of the compliance tensor,
$\alpha^{(n,m)}_{ij}$, $n \neq m$, in the elastic problem depend only on
the bulk values of the elastic constants and not on their values in the
anomalous shells around the rigid spheres, we will calculate the off--diagonal component of the
inverse capacitance tensor, $C^{-1}_{nm}, n \neq m$,
for this system of two spheres and check that it
does not depend on the values of the dielectric constant ($\bar{\epsilon}$)
near either of the  conducting spheres.

To compute the mutual capacitance of two conducting spheres one generally
employs the method of images to iteratively fix the boundary conditions
($\phi = \mbox{const}$) on each sphere in turn.  This procedure leads to
a convergent series for the capacitance tensor of two conducting spheres\cite{Smythe:68}.
We apply a similar technique.  To obtain just the component of the
two by two inverse capacitance matrix that we seek we will study
the problem where one sphere (say $n=1$) has a unit charge on it
and the other sphere (say $m=2$) is charge neutral.  The matrix
element in question ($C^{-1}_{12}$) is then simply the potential
of sphere two.  Furthermore, since we intend to show
that this component of the inverse capacitance tensor is independent
of the value of the inner dielectric constant ($\bar{\epsilon}$)
only in the limit that the sphere--sphere separation
($L$) is large compared to the both the sphere and cavity radii, we may
truncate the series generated by the method of reflections at the first
term.  The higher--order reflections will contribute corrections to our result
that are smaller by factors of $a/L$ or $b/L$.   We will return to the
issue of higher order corrections in $b/L$ due to subdominant terms in the
elastic displacement field and higher order reflections.

At the lowest order reflection we may ignore sphere two while we discuss the
free and polarization charge distribution on sphere one and its surrounding
cavity.  That distribution is equivalent to a unit charge at the center of
sphere one and two shells of bound, polarization charge. One shell is  at the
interface of the sphere and the inner dielectric ($r=a$) and the second
shell is  at the interface of the inner and outer dielectric ($r=b$).
These two polarization charge
densities, $\sigma_{\rm inner}$ and $\sigma_{\rm outer}$
respectively, are both spherically symmetric, and due to the neutrality
of the dielectric layer, $a < r < b$, we have the relation
\begin{equation}
\label{charge-relation}
\oint d \Omega a^2 \sigma_{\rm inner} +
\oint d \Omega b^2 \sigma_{\rm outer} = 0.
\end{equation}
Thus, at distances $r>b$, including the position of sphere two and its
surrounding pocket, the electric field due to this charge distribution is
simply that of a unit charge at the center of the sphere one.

We now consider the potential of sphere two in this electric field.
Because the spheres are conductors, the potential of sphere two is the
 same as the potential at its center.  That potential
is due to the linear superposition of the potential of a unit charge
a distance $L$ away (at the center of sphere one) and that of two spherical
shells of polarization charge  centered on sphere two.
One shell is at the interface between sphere two and the
medium with dielectric
constant $\bar{\epsilon}$ while second shell is at the interface between that
dielectric material and the bulk dielectric (with dielectric constant
$\epsilon$).  These surface charge distributions, unlike those of
sphere one, are not spherically symmetric.  However, the surface integrals
of the two polarization charge densities over the two surfaces vanish
independently of each other.  Thus the net effect on the potential at the
center of  sphere two due to each shell of polarization charge density
is zero.  The potential at the center of sphere two is then solely due
to the distant charge on sphere one; we find that the potential of sphere
two is simply $1/(4 \pi \epsilon L)$.  To lowest order in reflections we
have shown that
\begin{equation}
\label{mutual-cap}
C^{-1}_{12} = \frac{1}{4 \pi \epsilon L} + \cdots
\end{equation}
where the additional terms (not shown) come from higher order reflections.
These higher order reflections will generically depend on both $a,b$
and $\bar{\epsilon}$. A more detailed discussion of this derivation
in addition to a
discussion of the  form of the higher order reflections is given in
appendix \ref{higher-order-corrections}.

It is worth while at this stage to point out that to the same level of
approximation (lowest order reflections) the potential on sphere one due to
a unit charge on sphere one does, in fact, depend on the properties of
the inner, dielectric layer.  The potential on sphere one given by the
diagonal element of the inverse capacitance matrix is
\begin{equation}
\label{self-cap}
C^{-1}_{11} = \frac{4 \pi a b \bar{\epsilon}}{b + a \left(
\frac{\bar{\epsilon}}{\epsilon} - 1 \right) } + \cdots  \, \, .
\end{equation}
Once again, the additional terms not shown come from higher order
reflections.

Based on this simple analysis it seems reasonable to explore the
elastic problem in more detail to determine if this basic result
holds in the actual problem of rheological interest.

\section{The Viscoelastic Problem}
\label{visco}

\subsection{The homogeneous medium}
\label{visco-homo}

We begin the study of the viscoelastic problem by considering the displacement
field produced by the displacement of a rigid spherical particle embedded in
a homogeneous, elastic medium.  A sphere of radius $a$ is displaced by
$\epsilon \hat{z}$. We now calculate the resulting displacement field.

Local force balance in the medium demands that the displacement field obey the
partial differential equation
\begin{equation}
\label{force-balance}
0 = \mu \nabla^2 \bu + \left(\mu + \lambda \right) \bnab \bnab \cdot \bu
\end{equation}
where $\mu $ and $\lambda $ are the two Lam\`{e} constants characterizing the
isotropic, elastic medium.  Eq.~(\ref{force-balance}) is supplemented by
boundary conditions at the surface of the sphere and at infinity:
\begin{eqnarray}
\label{boundary-condition1}
\bu \left( \left| \bx \right| = a  \right) &=& \epsilon \hat{z}\\
\label{boundary-condition2}
\lim_{\left| \bx \right| \longrightarrow \infty} \bu \left(\bx \right) &=& 0.
\end{eqnarray}
Our solution of this problem is aided by two basic points: (i) the solution
must be azimuthally symmetric, and (ii) the solution must be linear in
$\epsilon \hat{z}$.  With their aid, we immediately write the
most general possible form of the displacement field
\begin{equation}
\label{general-sol}
\bu \left( \bx  \right) = \sum_{n} A_n \frac{ \hat{r} \cos \theta}{r^n} +
\sum_{m} B_m \frac{ \hat{z}}{r^m}.
\end{equation}
We have chosen these two sets of terms since they constitute the only solution
of Eq.~(\ref{force-balance}) that are azimuthally symmetric. As we will see,
$A_n$ and $B_n$ are proportional to $\epsilon$ so that ${\bf u}$ satisfies
the requirement of linearity in $\epsilon$.
We now put our ansatz, Eq.~(\ref{general-sol}), into the partial
differential equation, Eq.~(\ref{force-balance}).  Writing the radial
($\hat{r}$) and polar ($\hat{\theta}$) components of that expression
separately we find:
\begin{equation}
\label{matrix-sol}
{\cal M}(n,\zeta) \cdot \left( \begin{array}{c}
            A_n \\
            B_n
              \end{array} \right)
                        = 0
\end{equation}
where the $2 \times 2$ matrix $ {\cal M}(n,\zeta)$.
depends on the Lam\'{e} constants only through the
dimensionless ratio, $\zeta = \left( \mu + \lambda \right)/\mu $.  This matrix
is given by:
\begin{equation}
\left( \begin{array}{cc}
 \left[(1 + \zeta) (n-2) (n+1) -2\right]  & \left[ (1+\zeta) n (n+1) -2 n
\right] \\
(1 + \zeta) (n-2) -n & (1 + \zeta) n - n^2
\end{array} \right).
\end{equation}
A necessary and sufficient condition for Eq.~(\ref{matrix-sol}) to be
satisfied for nontrivial values of
$A_n$, $B_n$ is that
$\det {\cal M}(n,\zeta) = 0$.  There are four such solutions: $n = -2,0,1,3$.
By finding the eigenvectors associated with these eigenvalues we may write the
most general solution of Eq.~(\ref{force-balance}) consistent with the
two conditions discussed above,
\begin{eqnarray}
\label{ans}
\nonumber
\bu(\bx) &=& \frac{ a C_1 }{r} \left[ \gamma_1 \hat{r} \cos \theta +
\hat{z} \right] + \frac{ a^3 C_2 }{r^3} \left[ 3 \hat{r} \cos \theta -
\hat{z} \right] \\
  & & + C_3 \hat{z} + \frac{ C_4 r^2}{a^2}
\left[\gamma_2 \hat{r} \cos \theta - \hat{z} \right]
\end{eqnarray}
where the constants $C_m$ are determined from boundary conditions
 and the two dimensionless constants
\begin{eqnarray}
\label{gamma1}
\gamma_1 &=& \frac{1}{3 - 4 \sigma} \\
\label{gamma2}
\gamma_2 &=& 2 \left( \frac{ 2 - 3 \sigma}{3 - 2 \sigma} \right)
\end{eqnarray}
are functions of the Poisson ratio:
\begin{equation}
\label{Poisson_ratio}
\sigma = \frac{ 1}{2} \frac{\lambda }{\mu + \lambda }.
\end{equation}
Since the Poisson ratio can vary between $-1$ and
$1/2$\cite{Landau:Elasticity},
$1/7 < \gamma_1 < 1$ and $ 1/2 < \gamma_2 < 2$.  In the incompressible
limit ($\lambda \longrightarrow \infty$) $\sigma \longrightarrow 1/2$,
$\gamma_1 \longrightarrow 1$, and $\gamma_2 \longrightarrow 1/2$. Since we
are considering complex, frequency--dependent bulk and shear moduli, $\sigma
= \sigma(\omega)$ depends on frequency and is in general complex.

Examining the solution we see that the first term on the RHS of  Eq.~(\ref{ans})
decays only as $1/r$ away from the rigid sphere.  The second term is a dipole
field.  The third term is simply a constant shift of the entire medium which
is clearly a solution, but cannot contribute to the stress tensor.
The fourth term grows as $r^2$ as one moves away from the sphere.
In order to satisfy the boundary conditions at infinity for the problem under
consideration we must set $C_3 = C_4 = 0$.  The remaining two constants are
determined by the boundary conditions at the surface of the rigid sphere.  We
find
\begin{equation}
\label{answer-displacement-field}
\bu(\bx) = \frac{ \epsilon}{2} \left[ 3 \frac{ a}{r} \left( \eta_1 \hat{r}
\cos \theta + \eta_2 \hat{z} \right) - \eta_1 \frac{ a^3}{r^3} \left(
3 \hat{r} \cos \theta - \hat{z} \right) \right]
\end{equation}
where $\eta_1 = 1/(5 - 6 \sigma)$ and $\eta_2 = (3 - 4 \sigma)/(5 - 6 \sigma)$.
We note that in the incompressible limit, the displacement field around the
displaced sphere takes the form of what would be the perturbation
of the velocity
field of an incompressible fluid produced by the same sphere inserted in a
uniform flow in the $\hat{z}$ direction (at low Reynolds number).

Since we wish to calculate the response of the sphere to an applied force
we need to determine the force applied to the sphere that resulted in
the imposed
displacement of $\hat{z} \epsilon $.  To do this we calculate the restoring
force of the medium on the sphere.  The external force ${\bf F}$ is the
negative of the force the medium exerts on the sphere.  We can calculate the
latter force, which by symmetry must point in the $\hat{z}$--direction by
integrating the stress tensor over the surface of the sphere to obtain
\begin{equation}
\label{stress}
F_z = - \oint a^2 d \Omega \left[ \sigma_{rr} \cos \theta - \sigma_{r \theta}
\sin \theta \right],
\end{equation}
From this result we obtain the response
function\cite{Schnurr:97}
\begin{equation}
\label{answer-response}
\alpha(\omega) = \frac{\partial \epsilon }{\partial F_z} =
\frac{ 1}{6 \pi a \mu(\omega) }
\left[ 1 + \frac{ \sigma(\omega) - 1/2}{2 \left( \sigma(\omega) -1 \right) }
\right].
\end{equation}
We note that in the incompressible limit,
$\sigma(\omega) \longrightarrow 1/2$, we
recover the form of the Stokes mobility of the sphere in an
incompressible fluid.  The only difference between that result and
Eq.~(\ref{answer-response}) in the incompressible limit is the substitution of
the shear modulus, $\mu(\omega)$, for $ i \omega \eta$.

\subsection{The inhomogeneous medium: the results for distant particles}
\label{visco-inhomo}

Having solved the single--sphere problem, we are in a position to extend the
analysis to the two--sphere response function in a spatially
inhomogeneous elastic
medium.  As in the analogous electrostatic problem, we approach this problem via
the method of reflections.  To compute the response function to lowest order,
we simply need to calculate the displacement field at the location of the
second sphere due to a force applied to the first sphere. Once again, we model
the inhomogeneous elastic medium by the simple, anomalous pocket discussed
in our study of the analogous electrostatic problem.  We assume that the
spheres are surrounded by a spherical pocket
of material (of radius $b$) with elastic properties characterized
by the Lam\'{e} coefficients, $\bar{\lambda}, \bar{\mu}$.
See figure \ref{sphere-pic}.  The bulk material, far from the rigid spheres
has Lam\'{e} constants: $\lambda, \mu$.

Using Eq.~(\ref{ans})
we write down solutions to the force balance equations that apply in the
inner, anomalous region, and the outer bulk material respectively,
\begin{eqnarray}
\nonumber
\bu^{\rm i}(\bx) &=& \frac{a C^{\rm i}_1 }{r} \left[ \bar{\gamma}_1
\hat{r} \cos \theta +
\hat{z} \right] + \frac{ a^3 C^{\rm i}_2 }{r^3} \left[ 3 \hat{r}
\cos \theta - \hat{z} \right] +  \\
\label{u-inner}
  & & + C^{\rm i}_3 \hat{z} + \frac{ C^{\rm i}_4 r^2}{a^2}
\left[\bar{\gamma}_2 \hat{r} \cos \theta - \hat{z} \right] \\
\label{u-outer}
\bu^{\rm o}(\bx) &=& \frac{b C^{\rm o}_1 }{r} \left[ \gamma_1
\hat{r} \cos \theta + \hat{z} \right] + \frac{ b^3 C^{\rm o}_2 }{r^3}
\left[ 3 \hat{r} \cos \theta - \hat{z} \right].
\end{eqnarray}
In the above equation, $\bar{\gamma}_{1,2}$ are identical to the $\gamma$'s
defined in Eqs.~(\ref{gamma1})--(\ref{gamma2}) with the Poisson ratio
equal to that of the inner material.  Using the boundary condition at infinity,
Eq.~(\ref{boundary-condition2}), we have set $C_{3,4}^{\rm o} = 0$.  We are
left with six remaining constants that are determined by two boundary conditions
at the surface of the sphere [see Eq.~(\ref{boundary-condition1})] and four
boundary conditions at the interface of the two different elastic media, $\left|
\bx \right| = b$.  These four conditions enforce the continuity of the
displacement field: $\bu^{\rm i}(\left| \bx \right| = b) = \bu^{\rm o}
(\left| \bx \right| = b)$ and stress tensor: $ \sigma_{rj}^{\rm i}
(\left| \bx \right| = b ) = \sigma_{rj}^{\rm 0}(\left| \bx \right| = b)\, ,
j = r, \theta$ at that interface. These conditions are sufficient to determine
the six remaining constants.

Recall that we wish to show that the long--range part of the interparticle
response
function measures the bulk material properties of the medium independently
of the local modification of the material's elastic properties
by the rigid spheres.  In order to do this we first concentrate on the
part of the displacement field $\bu^{\rm o}(\bx)$ that varies as $1/r$.  We
will independently solve for the coefficient of this term.  Such a solution
allows a good approximation to the displacement field in the far--field regime
and will test the ideas discovered via the electrostatic analogy.

To calculate the coefficients $C_1^{\rm o,i}$ we employ a
global constraint on the
stress tensor: the integral of the flux of the stress tensor,
$\sigma_{ij} dS_j$, over any closed surface (with local outward normal
parallel to $dS_j$) enclosing the rigid sphere, which applies a force
$\bF$ to the elastic medium, must be equal and opposite to that applied force.
The integral of the stress tensor over such a surface is $-\bF$.
Thus we may write this
condition, for a particular spherical surface of radius $r$, with $r > a$
in the following form:
\begin{equation}
\label{surface-integral}
F_z = - \oint r^2 d \Omega \, \sigma^{\rm i,o}_{r z},
\end{equation}
where choice of the appropriate form of the stress tensor,
$\sigma^{\rm i}_{r z}$ or $\sigma^{\rm o}_{r z}$, is determined by magnitude
of $r$, {\it i.e.\/} whether the surface of integration is contained in the
inner region or in the bulk material.
In the above equation we have taken the force on the sphere to be in the
$\hat{z}$--direction and the integral is over all solid angles.  Counting
powers of $r$ in the stress tensor and noting that $\sigma \sim \nabla \bu$,
we find that only the part of the stress tensor coming from the term in $\bu$
proportional to  $C_1^{\rm o}$ can contribute to the result.  This term, which
depends on the radial distance from the sphere as $1/r^2$ is the only one
that will lead to an $r$--independent
result on the RHS of Eq.~(\ref{surface-integral}).  Since the LHS of this
equation is clearly $r$--independent, the other contributions to the stress
tensor coming from $C_n$, $n > 1$ must all vanish under the angular integration.

From the global stress constraint [Eq.~(\ref{surface-integral})] and our
solution for the displacement field we determine the coefficient
$C_1^{\rm o}$ to be
\begin{equation}
\label{1/r-part}
C_1^{\rm o} = \frac{ 1}{8 \pi a \mu } \left[ \frac{ \lambda + 3 \mu }{\lambda
+ 2 \mu } \right].
\end{equation}
The analogous coefficient in the inner region, $C_1^{\rm i}$, is given by
the same expression, however, the Lam\'{e} coefficients take the values of
the inner region: $\bar{\mu}, \bar{\lambda}$. We may use the above result to
eliminate one variable from the set of six that must be determined to completely
solve the present elastic problem. Before we continue this program, however, it is useful to calculate the far-field part of the viscoelastic, interparticle
response function. We have already seen, from the electrostatic analogy, that
only the dominant long-range part of the sphere--sphere interaction is expected
to be free of the influence of the anomalous pockets.
We seek, therefore,  to demonstrate, in a manner analogous to the
problem of the inverse capacitance of two conducting spheres in an inhomogeneous
dielectric, that the {\it inter\/}--particle response
function is independent of the rheological properties of the local pockets
surrounding the particles in the visco--elastic problem.

To do this we again use the lowest--order term in
the series solution of the two--sphere problem that is generated by the
method of reflections.  This lowest--order term simply gives the displacement of sphere 2 in response to an applied force on sphere 1 as the value of the
displacement field at the location of sphere 2 due to the displacement of
sphere 1, where that displacement field is calculated without regard to the
boundary conditions on sphere 2 or its surrounding shell of perturbed material.
Thus the solution of the far-field part of the single--sphere problem is
precisely the result that we need.  The corrections to this result coming
from higher--order reflections will be smaller than the previously calculated
part by a non-zero power of $b/r$, where $\br$ is the separation vector
between the two spheres.  These corrections are discussed in appendix
\ref{higher-order-corrections}.
Ignoring higher--order corrections coming from both
higher order reflections and the dipolar part of the far-field $\bu$,
we find
\begin{equation}
\label{response}
\alpha^{(21)}_{ij} = \alpha_{||}(r) \hat{r}_i \hat{r}_j + \alpha_{\perp}(r)
 \left(
\delta_{ij} -  \hat{r}_i \hat{r}_j \right),
\end{equation}
where the response along the line of centers is given by
\begin{equation}
\label{response-parallel}
\alpha_{||}(r) = \frac{ 1}{4 \pi r \mu(\omega)},
\end{equation}
and the response perpendicular to the line of centers is
\begin{equation}
\label{response-perpendicular}
\alpha_{\perp}(r) = \frac{ 1}{8 \pi r \mu(\omega)}
\left[ \frac{ \lambda(\omega) +  3 \mu(\omega)}{\lambda(\omega) + 2 \mu(\omega)}
\right].
\end{equation}
We have explicitly written the frequency dependence of the Lam\'{e} to
emphasize the applicability of this calculation to the complete viscoelastic
problem.  We note, however, that by neglecting inertial terms (which has
been justified previously in the single--sphere case at frequencies
of experimental interest\cite{Levine:2000}),  we are here imposing a
more stringent requirement.  The above result assumes that two spheres are
close enough that there is no significant phase shift between the oscillation
of the two spheres at the probing frequency, $\nu$
{\it i.e.\/} $\left| \br  \right| \ll c/\nu $, where $c$ is the speed of sound
in the medium.  Even for soft materials with relatively high compressibility,
it is possible to have
the necessary separation of length scales, $ b \ll r \ll c/\nu$
for Eq.~(\ref{response})-- (\ref{response-perpendicular}) to hold at
all experimentally accessible frequencies.

Finally it is interesting to
observe that in the incompressible limit, $\lambda(\omega) \longrightarrow
\infty$, the ratio of the response along the line of centers to that
perpendicular to the line of centers is $2:1$.  The experimental determination
of the deviation of this ratio from $2:1$ measures the compressibility of the
material at the frequency of observation.

\subsection{Single particle response in the composite medium}
\label{single-sphere}

It is interesting to compare the above results for the interparticle
response function in the composite (two-shell) medium with the
single--particle response in the same medium.  From the electrostatic analogy
we expect to find that the single--particle response function depends
on the elastic properties of both types of materials making up the
composite medium.  Below we will show this to be the case.  That
calculation also demonstrates that the comparision of the
single particle response to the two--particle response functions
allows one to determine the material properties of both materials
making up the composite medium.  This result shows,
at least within the simplified pocket model of the inhomogeneous medium, that
 measurements
of the probe particle autocorrelations combined with two--point
measurements of distant particles completely characterize
the bulk material and perturbation zone surrounding the probe.  In a
later section we will revisit this result and show that even in a more
physical model, in which the material properties of the medium vary
continuously with distance from the probe, it is still possible to
extract information about the perturbed region (as well as the
bulk properties) from a combination of one- and two--point correlation
measurements.

In order to solve for the single--particle response function, we must continue
along the lines of the previous section and solve for the complete deformation
field in the two--shell medium surrounding a particle.  As above we put a force
${\bf F} = F \hat{z}$ on the particle and determine the deformation field.
From the value of that field at the surface of the probe sphere
($\left| r \right| = a$)
we calculate the displacement of the probe and thus the response function
in question. Returning to Eqns.~(\ref{u-inner}),(\ref{u-outer}) we note that
there are now only four undetermined coefficients: From Eq.~(\ref{1/r-part}) we
already know $C^{\rm o}_1$, $C^{\rm i}_1$ in terms of the applied force ${\bf F}$.
We now continue with the simple but tedious task of matching boundary conditions
at the interface of the two elastic media and at the surface of the sphere as
discussed in the previous section.

At the surface of the sphere we find that:
\begin{eqnarray}
\label{r=a1}
C^{\rm i}_1 - C^{\rm i}_2 + C^{\rm i}_3 - C^{\rm i}_4 &=& \epsilon \\
\label{r=a2}
\bar{\gamma}_1 C^{\rm i}_1 + 3 C^{\rm i}_2 + \bar{\gamma}_2 C^{\rm i}_4 &=& 0,
\end{eqnarray}
where $\epsilon$ is the displacement of the sphere in the $\hat{z}$
direction.  The above set of equations actually contributes only one relation
among the remaining four
unknown coefficients since Eq.~(\ref{r=a1})
only exchanges one of these unknowns
for the, as yet, undetermined sphere displacement.  It is this
quantity, however, that we need to determine the response function.

From the continuity of the displacement field, ${\bf u}$, at the interface of
the two elastic media ($r=b$) we find two more relations:
\begin{eqnarray}
\label{r=b1}
C^{\rm o}_1 - C^{\rm o}_2 &=& \beta C^{\rm i}_1 -  \beta^3 C^{\rm i}_2
+  C^{\rm i}_3 - \beta^{-2} C^{\rm i}_4  \\
\label{r=b2}
\gamma_1 C^{\rm o}_1 + 3 C^{\rm o}_2 &=&  \beta \bar{\gamma}_1 C^{\rm i}_1
+ 3  \beta^3 C^{\rm i}_2 + \beta^{-2} \bar{\gamma}_2 C^{\rm i}_4,
\end{eqnarray}
where $\beta = b/a$.

For the remaining relation needed to specify all four undetermined
coefficients, we require the continuity of one component of the
stress tensor across the interface of the two media ($r=b$).  We
choose to consider
$\sigma_{r \theta} = \mu \left[ \partial_\theta u_r/r + \partial_r
u_\theta - u_\theta/r \right]$.  This yields the condition:
\begin{eqnarray}
\nonumber
\mu \left[ C^{\rm o}_1 \left( 1 - \gamma_1 \right) - 6 C^{\rm o}_2 \right] &=&
\bar{\mu} \left[ \beta  \left( 1 - \bar{\gamma}_1 \right) C^{\rm i}_1 +  \right.
\\
\label{r=b3}
 & & \left. - 6
\beta^3 C^{\rm i}_2 + \beta^{-2} \left(2 - \bar{\gamma}_2 \right) C^{\rm i}_4
\right].
\end{eqnarray}
We now have four equations to determine the unknown coefficients:
$C^{\rm i}_{2,3,4}, C^{\rm o}_2$ and another equation to eliminate one of
these four coefficients in favor of the quantity that we seek --
the displacement of the sphere, $\epsilon$.
The response function for the single sphere in the two--shell medium
is then given by
\begin{equation}
\label{single-particle-answer}
\frac{\epsilon}{F} = \alpha^{(1,1)}_{ij} =
\frac{1}{6 \pi \mu a} Z(\bar{\gamma}_1,\bar{\gamma}_2,\beta) \delta_{ij}.
\end{equation}
The response function has been
written as the product of the single particle response in an incompressible
bulk material with shear modulus
$\mu$ and a correction factor, $Z(\bar{\gamma}_1,\bar{\gamma}_2,\beta)$,
which depends on the ratio of the radius of the
anomalous pocket to the radius of the sphere, $\beta = b/a$,
and {\it all} of the elastic constants.  The correction
factor in terms of the constants $\gamma_{1,2}$ [defined in
Eqns.~(\ref{gamma1}), (\ref{gamma2})] is given by:
\begin{equation}
\label{correction-factor}
Z(\bar{\gamma}_1,\bar{\gamma}_2,\beta) = \frac{z_1}{z_2}
\end{equation}
where
\begin{eqnarray}
\label{z-top}
z_1 &=& 6 \beta^5 \left(\bar{\gamma}_1 + \bar{\gamma}_2 \right) \bar{p} \kappa
( \kappa - 1) + \\
\nonumber
 & & ( \bar{\gamma}_1 + 3) \bar{p} \kappa ( \bar{\gamma}_2 - 2 - 2
\bar{\gamma}_2 \kappa ) + 2 \beta^6 \bar{\gamma}_2 (\kappa - 1 ) \times \\
\nonumber
  &\times & \left(
p ( \gamma_1 + 3 ) -
 \kappa \bar{p} ( \bar{\gamma}_1 + 3 ) \right) +
\beta \left(  ( \gamma_1 + 3) ( \bar{\gamma}_2 - 2 ) p  + \right. \\
\nonumber
 & &  - ( 3 + 5 \bar{\gamma}_2
+ \gamma_1 ( 3 + \bar{\gamma}_2) ) p \kappa + \bar{p} \kappa  ( 9 - 4
\bar{\gamma}_2 - \bar{\gamma}_1 +  \\
\nonumber
& & 6 (\bar{\gamma}_1 + \bar{\gamma}_2 ) \kappa
) + \beta^3 ( \bar{\gamma_2} - 3) \kappa   \\
\nonumber
 &\times & ( ( - p ( \bar{\gamma}_1 + 1)
 \bar{p} ( 1 + \bar{\gamma}_1 ( 4 \kappa - 3 ) ) )
\end{eqnarray}
and
\begin{equation}
\label{z-bottom}
z_2 = 4 \left( \bar{\gamma}_2 \left[ 1 + 2 \beta^5 (\kappa - 1) - 2 \kappa \right]
\right),
\end{equation}
with $\kappa = \mu / \bar{\mu}$ and
\begin{equation}
\label{p-def}
p = \frac{\lambda + 3 \mu }{\lambda + 2 \mu};
\end{equation}
there is a corresponding term $\bar{p}$ which applies to the material of the
inner region.

It may be checked that the above expression
[Eqs.~(\ref{single-particle-answer}) -- (\ref{z-bottom})], reduces to
the simpler result for the single particle response function in a
homogeneous medium, Eq.~(\ref{answer-response}), when the elastic
properties of the two shells are equated.  As expected the full result
is a complicated function of both the elastic constants of the inner,
perturbed shell of the material, and the range of the perturbation: $b$.

Both because of the complexity of the above result and because many
applications of the these techniques apply to systems that are
essentially incompressible (polymeric solutions and melts fall
into this category) it is worthwhile to also record a simplier
version of the response function that obtains when both the
perturbed and the bulk material may be considered to be incompressible.
In that limit we find that the correction factor takes the form:
\begin{equation}
\label{single-particle-incom}
\frac{4 \beta^6 {\kappa'}^2 - 10 \beta^3 \kappa' + 9
\beta^5 \kappa' \kappa - 2 \kappa \kappa'' - 3 \beta
( 2 + \kappa  - 3 \kappa^2 )}{2 \left[ \kappa'' - 2 \beta^5 \kappa' \right] }
\end{equation}
where $\kappa' = \kappa -1$ and $\kappa'' = 3 + 2 \kappa$.

We end this section of the paper by noting that the above calculations
not only give the complete result for the single--particle response
function in the two material composite medium but they also determine
the next--to--leading order corrections for the interparticle
response function of two spheres in the same composite medium.  We discuss
this point further in Appendix \ref{higher-order-corrections}.  Here we
record the coefficient of the dipolar term in the displacement
field.  Based on arguments presented in Appendix
\ref{higher-order-corrections}, it can be shown that this dipolar term
gives the next--to--leading order correction in the two--particle response
function for distant particles.  The dipolar coefficient of the displacement
field in the bulk medium ($C^{\rm o}_2$) has been completely
determined already in the course of our solution of the single--particle
response function presented above.  We have found that in the case
where both media are incompressible it takes the form
\begin{equation}
\label{dipolar-coefficient}
C^{\rm o}_2 = - \frac{F}{8 \pi b \mu} \frac{ (1-\kappa) (3 + 2 \beta^5)
+ 5 \beta^2 \kappa}{3 ( 3 + 2 \kappa ) + 6 \beta^5 (1 - \kappa)}.
\end{equation}
As expected on more general grounds, this next--to--leading order correction
depends on all the elastic constants and the ratio of pocket radius to the
sphere radius.

\subsection{Differential Shell Method}
\label{differential-shell}

A more physical model of the anomalous region surrounding the probe particle
allows for the rheological properties of the medium to vary {\em continuously}
with distance from the probe.  In order to perform quantitive fits to the
single particle response function measured
in a complex fluid via microrheology, it is necessary to fit the data to
a continuous model of the anomalous zone.  As we will see, this fit requires
a theoretical model of the variation of the complex material's rheological
behavior as a function of distance from the sphere.  In this section, we
first present a general set of equations describing the variation of the
four displacement--field coefficients with distance for a given functional form
of the variation of the shear modulus with distance from the probe: $\mu =
\mu(\left| {\bf r} \right| ) $.  As an illustration of this method we then
apply our procedure to the case of a polymer solution at concentrations near
$c^\star$, the overlap concentration.

\begin{figure}[bp]
\scalebox{0.45}{\includegraphics{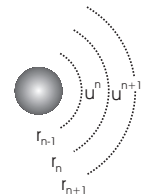}}
\caption{Differential shell method: Using stress and displacement continuity
at the interface of the $n^{\rm th}$ and $(n+1)^{\rm th}$ shells we determine
the coefficients of displacement field ${\bf u}^{n+1}$ in terms of the
coefficients in of the displacement field ${\bf u}^{n}$.  Later, taking the
thickness of the shells to zero: $\Delta r = r_{n+1} - r_n \longrightarrow 0$
we arrive at a set of differential equations governing the variation of
the displacement field coefficients -- the $C$s.
}
\label{shell-method}
\end{figure}

Having solved the two--shell model above, we can now generalize this technique
to many shells. To compute the displacement field in the continuous variation
limit, we divide the material into spherical shells of infinitesmal
thickness, $\Delta r$, centered on the probe particle.  Within each spherical
shell we may take the Lam\`{e} coefficients to be constant.
Now we can determine a relation between the set of displacement field
coefficients in the $n^{\rm th}$ shell ( $\left\{C_i^n \right\},
i = 1,\ldots,4$) to those
of the $n^{{\rm n}+1}$ shell ( $\left\{C_i^{{\rm n}+1}\right\},
i = 1,\ldots,4$) by using
the same set of boundary conditions at the interface of the two shells as
have been applied above. See figure \ref{shell-method}.
Taking the thickness of the shells to zero we
determine derivatives of the displacement field coefficients with respect to
$r$.  These linear differential equations can be integrated to give the
variation of the displacement field coefficients, and thereby determine the
form of the stain field.  The single--particle response function follows
naturally.  For the two--particle response function, we assume that there
still exists the large separation of length scales between the distance
separating the two probe particles and the distance over which there is
an appreciable variation of the elastic constants (the size
of the anomalous zone).The two--particle response function,
which does not depend on properties of the
anomalous zone, therefore, still applies.  Below we derive the differential
equations for the displacement field coefficients for the case where the
material is everywhere incompressible ($\lambda (r) = \infty$ for all $r$).

Matching the displacement field at the interface $\left| {\bf r} \right| = r_n$
we find two equations.  The first coming from matching the radial components
of the field is:
\begin{equation}
\label{diff-one}
\frac{ 2 a }{r_n} \Delta C_1 + \frac{ 2 a^3 }{r_n^3} \Delta C_2  + \Delta C_3
- \frac{ 1}{2} \left( \frac{ r_n}{a} \right)^2 \Delta C_4 = 0
\end{equation}
where $\Delta C_i = C_i^{n+1} - C_i^{n}$.  Taking the limit of thin shells {\em
i.e.}
\begin{equation}
\Delta C_i = \Delta r \, \left. \frac{ d C_i}{dr} \right|_{r=r_n} = \Delta r
\, \, \dot{C}_i,
\end{equation}
we arrive at the differential equation
\begin{equation}
\label{diff-two}
\frac{ 2 a }{r} \dot{C}_1 + \frac{ 2 a^3 }{r^3} \dot{ C}_2  + \dot{C}_3
- \frac{ 1}{2} \left( \frac{ r}{a} \right)^2 \dot{C}_4 = 0
\end{equation}
Using a similar procedure to match the $u_\theta$ parts of the displacement
field we find the differential equation:
\begin{equation}
\label{good-one}
\dot{C}_3 = - \frac{ a}{r} \dot{C}_1 + \left( \frac{ a}{r} \right)^3 \dot{C}_2
+ \left( \frac{ r}{a} \right)^2 \dot{C}_4.
\end{equation}

The remaining two equations come from matching the $\sigma_{r r} $ and
$\sigma_{r \theta}$ components of the stress tensor.  We recall from
section \ref{single-sphere} [see the discussion following
Eqns.~(\ref{r=a1})(\ref{r=a2})], however, that
only one more equation is needed to determine the coefficients since we
already know the functional form of $C_1$,
\begin{equation}
C_1 = \frac{1}{8 \pi a \mu(\omega)},
\end{equation}
from the same global property of
the stress tensor applied in the two--particle problem.  We again choose to
enforce the continuity of $\sigma_{r \theta}$, which, in the thin shell
limit, yields
\begin{equation}
6 \frac{ a^3}{r^4} \dot{C}_2 - \frac{ 3}{2} \frac{ r}{a^2} \dot{C}_4 +
 \frac{ \dot{\mu}}{\mu} \left[ 6 \frac{ a^3}{r^4} C_2 -
\frac{ 3}{2} \frac{ r}{a^2} C_4 \right]  = 0
\end{equation}

We now find two differential equations for the variables $C_2$ and $C_4$
by eliminating $C_3$ in Eq.~(\ref{diff-one}) using  Eq.~(\ref{diff-two}).
Furthermore we use our solution for $C_1$ to write the set of differential
equations in the following form [after undimensionalizing]:
\begin{eqnarray}
\label{B-one}
B_2'(x) + \frac{x^5}{6} B_4'(x) &=& - \frac{x^2}{3}
\frac{ d}{dx} \left( \frac{ \mu_0}{\mu(x)} \right)  \\
\label{B-two}
B_2'(x) - \frac{x^5}{4}  B_4'(x) &=& -  \frac{ d}{dx} \left( \ln \frac{ \mu}
{\mu_0} \right) \times \\
\nonumber
  & &  \left[ B_2(x) - \frac{ x^5}{4}  B_4 (x) \right]
\end{eqnarray}
where $x = r/a$, $\cdot '$ indicates a derivative with respect to $x$,
$\mu_0 $ is a modulus scale, $x = r/a$,  and
\begin{equation}
\label{B-def}
B_i = \frac{  8 \pi a \mu_0 C_i}{ F},
\end{equation}
where $F$ is the magnitude of the force applied to the sphere.  We find it
simpler, once again,  to study the response function by fixing a known
force on the sphere and computing its displacement, $\epsilon $.

Eqs.~(\ref{B-one}),(\ref{B-two}) can be integrated from the surface of
the probe sphere, $x = 1$.  Having a set of two first order differential
equations we require two boundary conditions to determine a unique solution.

There are two boundary conditions coming from the specification
of the displacement field at the surface of the sphere.
In general, this vectorial equation specifies two independent relations,
however, since the magnitude of the sphere's displacement is, as yet, unknown,
we obtain only one boundary condition for Eqs.~(\ref{B-one}) and (\ref{B-two}):
\begin{equation}
\label{B-bound-one}
B_2(1) + \frac{ 1}{6} B_4(1) = - \frac{ \mu_0}{3 \mu(x=1)}.
\end{equation}
The second equation
\begin{equation}
\label{B-bound-two}
\frac{ \mu_0}{ \mu(x=1)} - B_2(1) + B_3(1) - B_4(1) =  8 \pi \mu_0
a \frac{ \epsilon }{F}.
\end{equation}
coming from the boundary condition at the sphere expresses teh magnitude of
the sphere's displacement $\epsilon$, in terms of the $B$--amplitudes.
We still need another boundary condition to specify a unique solution of
Eqns.~(\ref{B-one}) and (\ref{B-two}).
The second boundary condition is
that $B_4$, the coefficient of the quadratically growing term in the
general solution of the displacement field with azimuthal symmetry, must
vanish in the large $r$ limit. Thus
\begin{equation}
\label{B-bound-three}
\lim_{x \longrightarrow \infty} B_4(x) = 0.
\end{equation}
Similarly, we know that there should be no constant term in the displacement
at large distances from the sphere so
$\lim_{x \longrightarrow \infty} B_3(x) = 0$. This boundary condition in
combination with Eq.~(\ref{good-one}) allows the determination of $B_3$ at
the surface of the sphere in terms of an integral over the (uniquely determined)
functions $B_2$ and $B_4$:
\begin{equation}
\label{find-B3}
B_3(1) = - \int_1^\infty \left\{ - \frac{ 1}{z} \frac{ d}{dz}
\left( \frac{ \mu_0}{ \mu(z)}\right)   + \frac{ 1}{z^3} B_2'(z) +
z^2 B_4'(z) \right\} dz
\end{equation}

The solution is effected by choosing $B_4(x=1)$ [using Eq.~(\ref{B-bound-one})
to determine $B_2(x=1)$] and then integrating the differential equations from
$x=1$ to infinity.  $B_4(x=1)$ is chosen so that this function
 goes to zero at large $x$.
Given this solution for $B_2$ and $B_4$ one can integrate
Eq.~(\ref{find-B3}) to determine $B_3(1)$.  Finally, with the full set of
initial values of $B_2,B_3$, and $B_4$ one can evaluate the response function
using Eq.~(\ref{B-bound-two}).

We further organize this calculation by defining the effective shear modulus
of the medium to be that value of the shear modulus needed to write the
response function  in the form that it would have taken in an
incompressible, {\em homogeneous} material.  In other words, we
define $\mu_{\rm eff}$ by
\begin{equation}
\label{def-mueff}
\frac{ \epsilon }{F} = \alpha = \frac{ 1}{6 \pi a \mu_{\rm eff} }.
\end{equation}
Here the vectorial indicies have been suppressed since, by rotational
symmetry, $\alpha_{ij}^{(1,1)}  \sim \delta_{ij}$ for an isolated sphere.
With  this definition we write the effective response function
in terms of the initial values of $B_2,B_3$, and $B_4$ and the modulus
scale as
\begin{equation}
\label{mueff}
\frac{ \mu_{\rm eff}}{\mu_0} = \frac{ 4}{3} \left[ \frac{ \mu_0}{\mu(1)} -
B_2(1) + B_3(1) - B_4(1) \right]^{-1}.
\end{equation}
We note that in the homogeneous medium: $B_3 = B_4 = 0$ and
$\mu_0/\mu(1) = 1$.  In addition, we find that $B_2 = -1/3$ so that $\mu_{\rm
eff} = \mu_0$ as required for consistency.

\begin{figure}[bp]
\scalebox{0.35}{\includegraphics{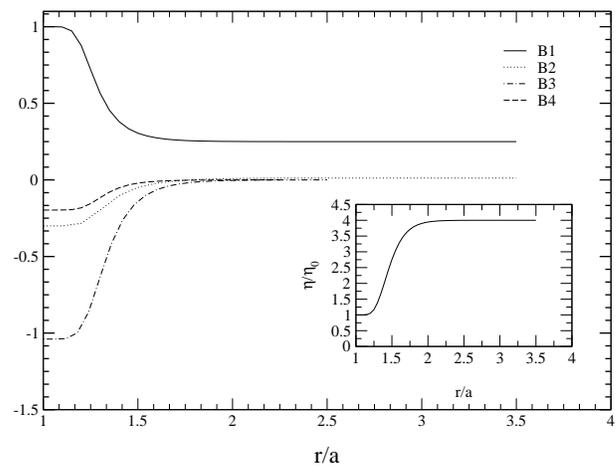}}
\caption{%
The variation of the dimensionless displacement field coeffecients computed
numerically.  The characteristic length scale for the variation of the
polymer concentration (thus the solution viscosity) is $0.30$ in the
dimensionless units $r/a$. The inset shows the variation of the
fluid viscosity with distance from the sphere.
}
\label{B-plot}
\end{figure}

As an example of the differential shell method we  consider the
case of a semi-dilute polymer solution -- see appendix~\ref{polymer-depletion}.
To apply the methods of one--point microrheology to this case,
one would measure the fluctuating position of a probe particle
in the liquid (due to Brownian diffusion) and compute from the position
autocorrelations the diffusivity of that probe.  Using the Stokes--Einstein
relation one could then extract a measurement of the viscosity.  However,
such a measurement should be an underestimate since the
probe sphere produces a spherical pocket of polymer--depleted solution
surrounding it (see Appendix~\ref{polymer-depletion} for details).  The
local polymer concentration will approach its bulk value essentially
exponentially  with distance from the sphere with a ``healing length''
controlled by the polymer correlation length in the solution.  This
polymer--depleted shell of fluid has a lower viscosity than that of the bulk.
We take a continuous polymer concentration profile suggested by
self-consistent calculations\cite{deGennes:79,worry} and numerically
integrate the differential equations for the case that
the polymer correlation length is 30\% of the sphere radius and the
bulk solution visocity is four times the value of that of the solvent.
The variation of the coefficients $B_1, \ldots,B_4$ with distance from the
probe sphere are shown in figure~\ref{B-plot}.

Using Eq.~(\ref{mueff}) we find that indeed the {\it single\/} particle
measurements suggest that the viscosity is smaller than it's bulk value.  For
this particular case the effect is small -- the viscosity measurement
coming from one--point microrheology is about 73\% of the actual
bulk value.  If the depletion zone were larger compared to the sphere radius,
the effect of the anomalously small value of the solution viscosity in the
depletion zone would be more significant.


\section{Summary}
\label{final}

In this paper we have studied the single--particle and two--particle response
functions in an inhomogeneous viscoelastic medium.
These response functions must be known in order to use
microrheological measurements as a probe of the material properties of soft
materials.  We restricted our analysis
to the type of inhomogeneity that is caused by the introduction of the
probe particles themselves. We have assumed that the rheological anomally in
the material relaxes to the unperturbed, bulk value as some function of radial
distance from the probe particles.  To make a model system that
has the simplest possible inhomogeneity of this form we have considered
not only the``pocket model'' (consisting of a spherical cavity surrounding each
probe particle with perturbed viscoelastic properties), but also a more physical
model in which the material's rheological properties vary continuously with
distance from the probe sphere.  We have also shown that the combination of
one--point techniques (which measure a combination of the properties of the
unperturbed, bulk material and the rheological anomalous material immediately
surrounding the probe sphere) and two--point techniques (which measures the
bulk rheological properties) allows the experimentalist to probe details of the
probe--particle, medium interaction.

We thank J.~C.~Crocker for many helpful discussions and communicating
unpublished data. We also thank
F. Pincus for a careful reading of the manuscript and  AJL
thanks Michael Cohen for useful
discussions. This work was supported in part by the NSF MRSEC program under
grant number OMR 00--79909.

\appendix

\section{Corrections for closer particles: Higher order reflections and
subdominant terms in the strain field}
\label{higher-order-corrections}

There are two classes of corrections to the result of section
\ref{visco-inhomo} for
the interparticle response function of two distant spheres.  These
corrections produce terms that are higher order in $a/R$ where $R$ is the
(large) separation of the two probe spheres.  In general, these corrections
depend on all the elastic constants of the composite medium.  It is
therefore important to at least estimate the relative importance of these
terms in order to determine how distant two probe particles must be in
order for their correlated fluctuations to be governed primarily by the
bulk elastic constants.

The two classes of corrections are due to
either subdominant terms in the displacement field of sphere one at the level
of the zeroth order reflection (in which we
ignore the role of second sphere in determining its subsequent displacement)
or corrections to the displacement field that
result from higher order reflections (iteratively correcting the boundary
conditions of $\bu$ at the surface of each sphere--and--pocket).
In this section we determine which of these effects first presents
deviations to the far--field results presented earlier.  We have already seen
that subdominant corrections in the far--field ${\bf u}$ are of a dipolar form,
decaying with distance as $R^{-3}$. These corrections also depend on the
properties of the inner pockets.  We now look at the corrections coming from
higher order reflections.

\begin{figure}
\scalebox{0.45}{\includegraphics{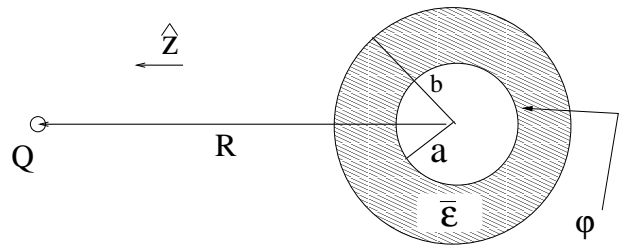}}
\vspace{0.1cm}
\caption{%
Schematic diagram of the simpler electrostatic problem designed to test
the importance of first order reflections upon the elastic response function.
The first sphere, that is charged, can be replaced by a simple point charge
to this order in the reflections.  We focus on the response (potential) of
the second sphere which sits in its pocket  of material with dielectric
\protect{ $\bar{\epsilon}$}.  The bulk material has dielectric constant
$\epsilon = 1$.}
\label{electro-fig}
\end{figure}

Because the
full elastic problem in the composite medium is quite complex, once again it is
helpful return to the electrostatic analogy for guidance.
As before we replace the rigid particle of
radius $a$ and its surrounding spherical pocket (of radius $b$)
of anomalous material by its electrostatic analog: a conducting sphere of
radius $a$ surrounded by a region of radius $b$ with dielectric
coefficient $\bar{\epsilon}$.  To simplify the formulae we set the
bulk dielectric constant to unity.
We consider first the potential at sphere two to lowest order.
At this iteratation we may still replace (the charged) sphere one
and its surrounding dielectric pocket by a point charge $Q$ at the origin of
that sphere.  First we calculate the potential at the second (uncharged) sphere
and then we determine the correction to that potential coming from higher
order reflections. See figure \ref{electro-fig} for
a diagram of the electrostatic problem under consideration.

Using the azimuthal symmetry of the problem we can write the general form
for the electrostatic potential, $\phi_{\rm outer}$, in the bulk material
($r > b$) by
\begin{eqnarray}
\nonumber
\phi_{\rm outer}(\br) &=&
\frac{ Q}{4 \pi } \sum_{\ell = 0}^\infty \frac{ %
r_<^\ell}{r_>^\ell} P_\ell (\cos\theta) + \\
\label{outer-pot}
 & & + \frac{ 1}{4 \pi \epsilon}
\sum_{\ell = 0}^\infty D_\ell r^{- (\ell + 1)} P_\ell (\cos\theta),
\end{eqnarray}
where $r_<$, $r_>$ are the minimum, maximum  of $r$ and $R$ respectively. The
functions $P_\ell(x)$ are the the standard Legendre polynomials.  The first
term represents the potential due to the point charge at $\br = \hat{z} R$ (
we take the origin of the coordinate system to be at the center of the sphere.)
and the second term gives the corrections to that potential field in the
bulk due to polarization charges induced at the interface of the two dielectric
media and on the conducting sphere.  These corrections are given in terms
of the yet unspecified coefficients
$D_\ell$.  We can similarly write the expression for the potential in the
pocket ($ a < r < b$),
\begin{equation}
\label{inner-pot}
\phi_{\rm inner} (\br) = \sum_{\ell = 0}^\infty \left[ E_\ell r^\ell +
F_\ell r^{- (\ell +1)} \right] P_\ell (\cos\theta),
\end{equation}
in terms of two sets of unknown coefficients, $E_\ell$  and $F_\ell$.

We are trying to find the potential of the conducting sphere two ($ \left| \br
\right| = a$).  Since the sphere is an equipotential surface, we find from
Eq.~(\ref{inner-pot}) that $E_\ell = F_\ell = 0$ for all $\ell \neq 0$.
Furthermore, if we define $\phi_0 $ to be the as yet unknown potential of the
sphere, we find that
\begin{equation}
\label{voltage}
E_0 + \frac{ F_0}{a} = \phi_0
\end{equation}
We now note that there are no free charges in the system other than the
distant charge $Q$, so the surface integral of the electric displacement
field over a sphere of radius $r$, $ b < r < R$ must vanish.  This condition
forces $D_0 = 0$.  From that result and the continuity of the radial component
of the electric displacement vector at the interface of the two dielectric
media ($r= b$) we determine that $F_0 = 0$ as well.  Finally, from the
continuity of tangential components of the electric field at the same
interface, ${\bf \nabla }  (\phi_{\rm inner}  - \phi_{\rm outer} )
|_{r = b} \times \hat{r} = 0$, we find that the remaining coefficient of
interest, $E_0$ is given by: $E_0 = Q/(4 \pi R)$ so from
Eq.~(\ref{voltage}) we arrive at the result
\begin{equation}
\label{volt-ans}
\phi_0 = \frac{ Q}{4 \pi \epsilon R}.
\end{equation}
This rather remarkable conclusion is that the presence of the anomalous
dielectric pocket does not effect the result of the zeroth order reflection
at all.  We can physically understand this result along the lines
presented in the text. We also note that the argument presented there can be
extended to an arbitrary number of such dielectric shells; this suggests that
the effect of {\it any\/} radially symmetric dielectric coefficient variation
on the potential of the central conducting sphere always vanishes at the
level of the zeroth reflection.

Next we consider the first correction to the potential on sphere two coming
from higher order reflections.  Since there is no free charge on either
sphere two, or its surrounding dielectric, we know that the integral of the
normal component of the electric displacement field over a surface just
inside the dielectric shell (and just outside the surface of sphere two)
vanishes:
\begin{equation}
\label{reflec1}
\oint_{r=a+}  D_r r^2 d \Omega = 0 \longrightarrow
\oint_{r=a^+}  E_r r^2 d \Omega = 0.
\end{equation}
The vanishing of the same integral of the normal component of the
electric field follows from the fact that the dielectric constant in
the material outside sphere two is assumed to be spherically symmetric.
This assumption is clearly valid in the somewhat artifical two--shell
model of the composite medium, but it should remain valid in a more
physical model in which the dielectric constants vary continuously
with distance from the probe particles, at least as long as the two
particles are farther apart that a ``healing length'' over which the
material recovers it bulk properties away from the probe particles.

We now study the polarization charge induced on the dielectric interfaces
surrouding sphere two. It is clear that we can replace the dielectric
shell around sphere two with two spherical surfaces of bound, polarization
charge density (at $r = a$ and $r=b$).  Our solution of the electrostatic
problem defined in figure \ref{electro-fig} shows that the bound polarization
charge on the outer surface of the pocket surrounding sphere two (at the
order of lowest reflections) is given by:
\begin{equation}
\label{s-outer}
\sigma_{\rm outer} (\theta) = Q \sum_{\ell = 1}^\infty \frac{b^{\ell -1}}
{R^{\ell+1}} \left( \frac{ 2 \ell +1}{\Gamma_\ell + \ell + 1} \right)
\left[ \frac{ 1 - \bar{\epsilon}}{\bar{\epsilon}} \right] P_\ell (\cos \theta)
\end{equation}
In the above equation, we have defined
\begin{equation}
\label{g-definition}
\Gamma_\ell = \bar{\epsilon} \frac{ \ell + (\ell +1)
\left( \frac{ a}{b} \right)^{2 \ell + 1}}{1 -
\left( \frac{ a}{b} \right)^{2 \ell + 1}}
\end{equation}
It is clear in the above result that if there is no dielectric discontinutiy
at the edge of the pocket ($\bar{\epsilon} = 1$) this polarization charge
density vanishes.

We now compute the bound charge at the interface between the conducting
sphere (two)
and the inner dielectric.  To distinguish the bound polarization charge from
that free charge on the conductor, we need to calculate the difference in
charge density at $r=a$ between the general case and the particular case
of no anomalous dielectric, {\it i.e.\/} we determine:
$\tilde{\sigma}(\theta ) = \sigma_{\rm inner}(
\theta ) - \sigma_{\rm inner}( \theta )|_{\bar{\epsilon } = 1}$ to be
\begin{equation}
\label{s-inner}
\tilde{\sigma}(\theta ) = Q \sum_{\ell = 1}^\infty  (2 \ell +1) \frac{ a^{
\ell -1}}{R^{\ell + 1}} \Xi_\ell  P_\ell(\cos \theta),
\end{equation}
where $\Xi_{\ell}$ is given by
\begin{equation}
\Xi_{\ell} =
\left\{ 1 - \frac{ 2 \ell + 1}{1 + \ell (\bar{\epsilon}
+ 1) - (1 - \bar{\epsilon}) (\ell + 1) \left( \frac{a}{b} \right)^{2 \ell + 1}
}
\right\}.
\end{equation}

Since we are interested in understanding the next-to-leading
order correction to the result presented in this article for the mutual
response function of distant spheres, we may approximate the three charge
distributions in the large $R$ limit as follows:  From the shell of bound
polarization charge at the outer interface surrounding sphere two ($r = b$)
we get an effective dipole moment of
\begin{equation}
\label{out-dipole}
{\bf P}_{\rm outer} =  - \frac{ 3 Q }{R^2} ( \bar{\epsilon}-1) b^3
\frac{ 1 + 2 \left( \frac{ a}{b} \right)^3}{2 + \bar{\epsilon} +2 (
\bar{\epsilon} - 1) \left( \frac{ a}{b} \right)^3} \hat{z}.
\end{equation}
From the shell of bound polarization charge on the inner interface of
the dielectric shell ($r=a$) we get the effective dipole moment
\begin{equation}
\label{in-dipole}
{\bf P}_{\rm inner} =   \frac{ 3 Q }{R^2} a^3
\left\{ 1 - \frac{ 3}{2 + \bar{\epsilon} +2 (
\bar{\epsilon} - 1) \left( \frac{ a}{b} \right)^3}
\right\} \hat{z}.
\end{equation}
Note for $\bar{\epsilon}  > 1$  The dipole moment of the outer shell points
away from the first sphere and the dipole moment of the inner shell is
anti-parallel to the outer dipole moment.  This shows that, back at the first
sphere, the net effect of these two effective dipoles is reduced by their
partial cancellation. Finally we include the effect of the polarization of the
conducting sphere two.  This is simply given by the standard answer from
the first reflection term for two spheres.  The surface charge distribution
of the conducting sphere produces the same field as a pair of point charges
of equal and opposite magnitude within the sphere: a charge of -$ Q a/R$
displaced from the center of sphere two towards the center of sphere one
by a distance of $ a^2 / R$ and the opposite charge
(to ensure the charge neutrality of sphere
two) at center of sphere two.  This charge distribution at large distances,
produces another dipolar field with dipole moment: $P_{\rm sphere} = -Q a^3/R^2
\hat{z}$.

The potential in the vicinity of sphere one
produced by the dipoles induced in the neighborhood of sphere two
is thus the sum of three dipole potentials,
each centered at the origin of sphere two.  Since the electric potential and,
 hence the electric field, is linear in the dipole moments we can approximate
the net electric field at sphere one as the field of a single dipole
located at the center of sphere two having a net dipole moment of:
\begin{equation}
{\bf P }_{\rm net} = {\bf P}_{\rm inner} + {\bf P}_{\rm outer} + {\bf P}_{\rm sphere}
\end{equation}
Collecting our previous results we find that the net dipole is given by:
\begin{equation}
{\bf P }_{\rm net} =- \hat{z} \frac{Q b^3}{R^2}  \left[ \rho^3 +
\frac{ \bar{\epsilon}-1}{2 + \bar{\epsilon} +2 (\bar{\epsilon} - 1)
\rho^3} \left( 1 - \rho^3 \right)
 \left( 1 + 2 \rho^3 \right) \right],
\end{equation}
where $\rho = a/b$.  The most significant point coming from the
calculation is that we have confirmed that the polarizibility of the
combination of the conducting sphere and dielectric shell does, in fact,
depend on properties of that dielectric shell.  Without performing any further
detailed calculations, we may assume that the polarizibility of the conducting
sphere plus dielectric pocket takes the
form: $ \alpha_p = \alpha_p (a/b,\bar{\epsilon}) b^3$.  The dipole moment
induced on sphere one by the dipole moment on sphere two then
has a magnitude of:
\begin{equation}
P_{1} = \alpha_p (a/b,\bar{\epsilon}) b^3  \times \frac{ b^3}{R^5} P_{\rm net}.
\end{equation}
In the above equation the first term in the product is the polarizibility of
sphere one and the second term is the electric field at sphere one due to
the polarization of sphere two.
So the shift in the potential of sphere two due to the next order reflection
must take the form:
\begin{equation}
\Delta \phi_2 \simeq P_1  \frac{ \alpha_p (a/b,\bar{\epsilon}) b^3}{R^2}
= \frac{ b^6 Q \alpha_p^2 (a/b,\bar{\epsilon})}{R^7}.
\end{equation}

Based on the electrostatic analog to the viscoelastic response function,
we see that the next--to-leading order term in the approximate solution for
the potential of sphere two decays as the seventh power of the sphere--sphere
separation.  The detailed calculation of the polarization sphere two serves
to confirm that all these higher order terms necessarily involve all the
properties of the anomalous pockets.
The principal point of this section remains that we can  conclude
that the subdominant term in the displacement
field in the elastic problem, which decays only as $R^{-3}$, give the
next-to-leading order correction for the interparticle response function.

\section{The viscosity of semi--dilute polymer solutions near the probe}
\label{polymer-depletion}

In the semi-dilute regime the polymer volume fraction, $\phi$,
lies in the range: $ \phi^\star \ll \phi  \ll 1$, where $\phi^\star$ is the
volume fraction at which the individual coils overlap.  Here we may approximate
the relaxational dynamics of a single chain as the reptation of a string of
blobs with mean radius equal to the polymer correlation length $\xi$ and thus
composed of
\begin{equation}
\label{g-def}
g = \left(\frac{\xi}{\ell} \right)^{5/3}
\end{equation}
monomers, where $\ell$ is the Kuhn length.  A polymer of $N$ monomers consists
of $N/g$ blobs, and its reptation time is
\begin{equation}
\tau_{\mbox{rep}} = \tau_{{\tiny \mbox{Zimm}}}(g)
\left( \frac{N}{g} \right)^3
\end{equation}
where $\tau_{\mbox{Zimm}} \sim \xi ^2/D(\xi )$ where $D(\xi ) \sim 1/\xi  $, the
diffusion constant of a sphere of diameter $\xi $.
is the Zimm relaxation time of a blob.  In the
semi--dilute regime, $\xi$ scales with polymer volume fraction as
\begin{equation}
\label{xi-scaling}
\xi = \ell \phi^{-3/4},
\end{equation}
which implies
\begin{equation}
g \sim \phi^{-5/4}.
\end{equation}
The reptation time thus scales with volume fraction as
\begin{equation}
\tau_{\mbox{rep}} \sim \phi^{3/2}.
\end{equation}
To find the contribution of $\tau_{\mbox{rep}}$ to the viscosity, we note
that $\eta_{\rm P} \sim G_0 \tau_{\mbox{rep}}$, where $G_0 \sim k_{\rm B} T/
\xi^3 \sim ( k_{\rm B} T/a^3 ) \phi^{9/4}$ is the plateau modulus of the
semi--dilute solution.  Thus
\begin{equation}
\eta_{\rm P} = \eta_0 \phi^{15/4}
\end{equation}
where $\eta_0$ is a viscosity.

Finally, in order to discuss the variation of the effective solution
viscosity near the surface of the probe sphere we need to understand the
polymer concentration profile near an impenetreble obstruction.  If we
assume that the correlation length in the solution is much smaller than
the radius of curvature of the probe sphere,  we may approximate the
polymer concentration profile surrounding the sphere by that of the profile
near a flat, hard wall.  This problem has been studied using self-consistent
methods with ground state dominance\cite{Moore:77,deGennes:79}.  The
solution for the concentration profile near a wall at $x=0$ is
\begin{equation}
\label{concentration-profile}
c(x) = c_0 \tanh^2 \left( \frac{ x}{\xi } \right),
\end{equation}
where $c_0$ is the bulk polymer concentration.  The correlation length
obtained from this calculation is known not to scale correctly with polymer
concentration; we suppliment the above solution with the correct scaling form
from Eq.~(\ref{xi-scaling}).  We also point out that if $\xi $ is compariable
to the sphere radius, the detailed form of Eq.~(\ref{concentration-profile})
 must be quantitively inexact.  The qualitive results of this analyis still
hold.  In particular, even for a sphere size that is comparable to the
correlation length, we expect that the recovery of the bulk viscosity occurs
over the length scale $\xi $ as one moves away from the sphere.


\begin{thebibliography}{worry}
\bibitem{worry}
It should be pointed out that the self-consistent calculation is not
quantitively correct. However, the existence of a mesoscopic length scale
in the polymer solution which is on the the order of the bead size is
clearly valid.  Given this one length scale, the essentially  expoenential
dependence of the polymer concentration around the bead is at least
reasonable.  We do not include the effect of the curvature of the bead's
surface in this calculation.
\end{thebibliography}
\end{document}